\documentclass[conference]{IEEEtran}

\IEEEoverridecommandlockouts

\usepackage{cite}

\ifCLASSINFOpdf
  \usepackage[pdftex]{graphicx}
  \DeclareGraphicsExtensions{.pdf,.jpeg,.png}
\fi
\usepackage[cmex10]{amsmath}
\usepackage{algorithmic}

\usepackage{array}
\begin{document}
\title{The Cloverleaf Antenna: A Compact Wide-bandwidth Dual-polarization Feed for CHIME}
\author{\IEEEauthorblockN{Meiling Deng$^a$ and Ducan Campbell-Wilson$^b$ for the CHIME collaboration}
\IEEEauthorblockA{$^a$Department of Physics and Astronomy,
The University of British Columbia,
Email: mdeng@phas.ubc.ca\\
$^b$School of Physics, The University of Sydney, Email: dcw@physics.usyd.edu.au}}
\maketitle
\IEEEpeerreviewmaketitle
\begin{abstract}
We have developed a compact, wide-bandwidth, dual-polarization cloverleaf-shaped antenna to feed the CHIME radio telescope. The antenna has been tuned using CST to have smaller than -10dB s11 for over an octave of bandwidth, covering the full CHIME band from 400MHz to 800MHz and this performance has been confirmed by measurement.  The antennas are made of conventional low loss circuit boards and can be mass produced economically, which is important because CHIME requires 1280 feeds. They are compact enough to be placed 30cm apart in a linear array at any azimuthal rotation.
\end{abstract}
\parskip=0.5pt

\section{Introduction}
We have built a novel, cloverleaf shaped compact dual-polarization
feed for the Canadian Hydrogen
Intensity-Mapping Experiment\cite{CHIME}.  CHIME is a radio telescope designed to
measure Baryon Acoustic Oscillations (BAO) by measuring the intensity of
neutral hydrogen over half the sky through the redshift range $0.8\leq
z\leq 2.5$.  At these redshifts the 21cm line of neutral hydrogen
appears in the frequency range 400MHz to 800MHz.
CHIME has no moving parts; it consists of five parallel cylindrical
parabolic reflectors, each 20m wide, 100m long and f/0.25.  Feeds
are spaced 30cm apart along each focal line. Signals are amplified and
brought to a single custom digital correlator.

The full instrument requires 1280 dual polarization feeds with an acceptable beam pattern, low material loss and s11 lower than -10dB from 400MHz to 800MHz.  With this many feeds, it is important that uniform, reliable feeds can be manufactured economically. Other solutions considered as CHIME feeds are the four-square antenna\cite{Martin}\cite{4square} developed for the Molonglo Telescope\cite{Molonglo}
, the four-point antenna\cite{4point} and the four-point antenna with tuning plate\cite{tuning}. All these feeds generate an approximately circular beam
suitable for feeding deep paraboloidal reflectors. The performance of these feeds
differs mostly in their matching bandwidth.
\section{Radiation mechanism}
\begin{figure}[!t]
\centering
\includegraphics[width=0.5\textwidth]{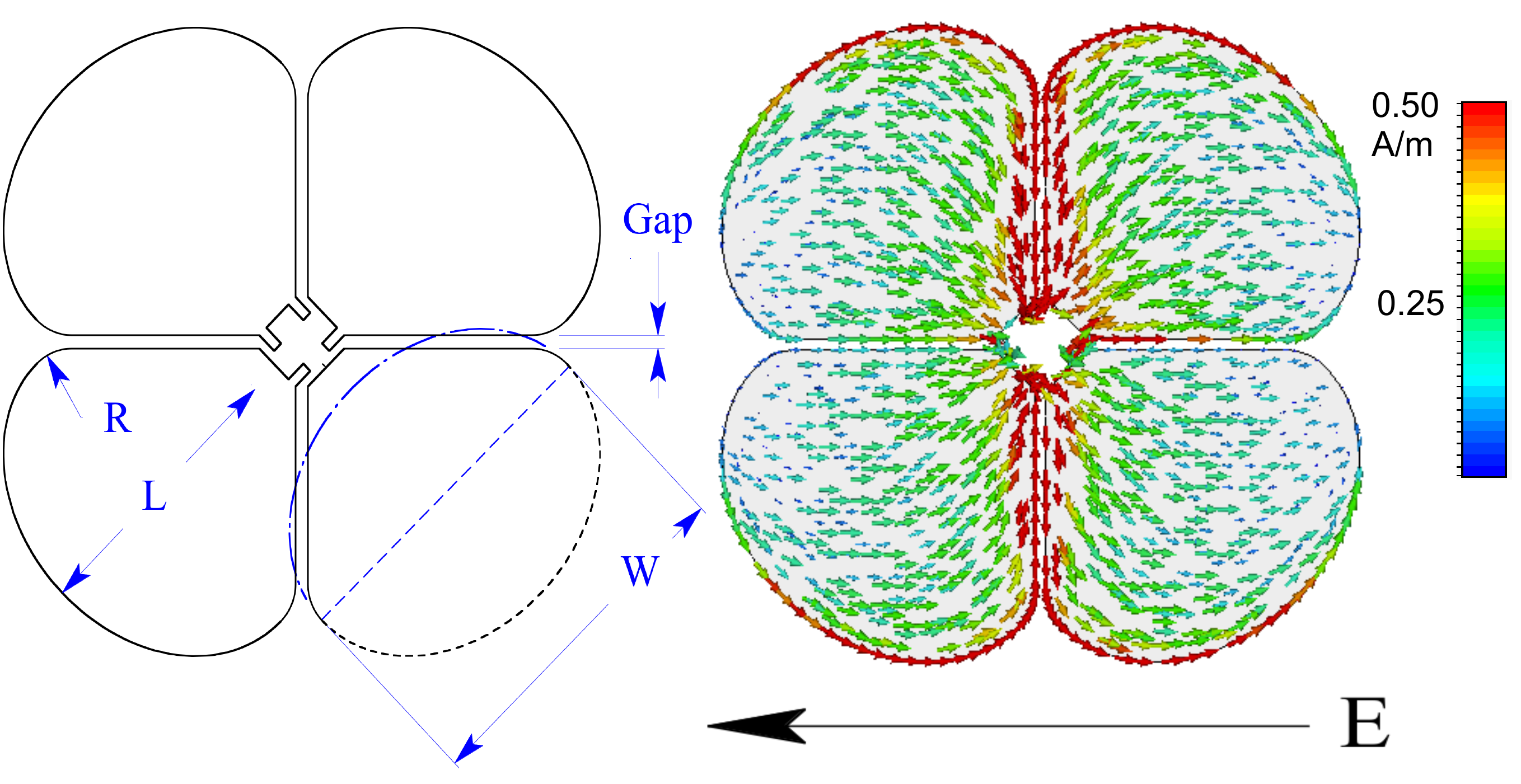}
\caption{At left, the shape of each petal 
consists of two perpendicular straight lines, two 45 degree circular arcs with radius
$R$ and one half an ellipse.  $W$ is the major axis of the ellipse and
$L$ is the length from the intersection of the straight sides to the outer
edge of the ellipse.  The shape is illustrated here for the adopted
values of gap, $R, L,$ and $W$. Each of the four tabs shown at the centre is
connected to one side of a vertical microstrip transmission line and in each
case the full width of the adjacent petal is connected to the other
lead. At right, CST simulated currents for one linear polarization at 600MHz are shown.  
Note the small asymmetry in the curent
distribution near the centre becasue of the tab geometry.}
\end{figure}
Our feeds are a modification of four-square antennas developed for the Molonglo Observatory.  The petals, stem and base are all made from printed circuits boards (PCB).  To broaden the bandwidth, we have modified the petals to have curved outer edges as shown at left in Figure 1, eliminating the depedence on a single dimension. The curves
are smooth and each petal is symmetric. CST simulated current pattern is shown at right in Figure 1 for one
linear polarization at 600MHz. The currents near the gaps between petals run in opposing directions so they cancel, and do not contribute to the radiation pattern.
For this polarization, farfield radiation arises from the coherent
currents running along the curved outer edges of the top and bottom pair of
petals. 
For each linear polarization, two differential signals, each from a pair of
petals, are combined through tuned baluns to form one single-ended
output. Thus each single polarization signal involves
currents in all four petals. This is called in-pair feeding.
Full baluns, from both polarizations, consist of four identical microstrip transmission lines 
along four vertical support boards(stem) and a horizontal base board. Both of the single-ended outputs are on the base board. Each transmission line is varied in several abrupt steps, and the lengths
and characteristic impedances of the transmission line segments are
tuneable. We have demonstrated that electrical losses in conventional
circuit board materials generate unacceptable noise levels for
astronomical instrumentation.  Teflon-based PCB is used everywhere
there is a transmission line. 
\IEEEpubidadjcol
\section{Tuning the antenna performance}
In order to tune the antenna parameters to produce acceptable
performance we have constructed a full CST
model of a cloverleaf antenna(only one polarization present).  To verify the procedure, we first built two different cloverleaf antennas with arbitrarily chosen shapes, measured their s11 and compared these measurements to CST simulations.  The comparison proves our CST simulation is reliable.

Measurements show that coupling between two polarizations and coupling between adjacent antennas do not affect s11. Therefore, we proceeded to iterate the cloverleaf design using CST following the plan listed below.

We initially fixed the parameters of the transmission lines to a design chosen
for ease of manufacture: the transmission line has two characteristic
impedances, one on the vertical support board and the other on the
horizontal baseboard.

We set the initial petal parameters to be $(R,W,L)=80, 140, 150)$mm
and altered $R, W,$ and $L$ successively, to learn which parameters have
the most impact on the antenna's s11. Altering $R$ has very
little impact, and we fixed it to $R=20$mm, the peak of a very shallow
performance curve.

We used the optimization algorithm implemented in CST to explore
s11 in $(W,L)$ space. Varying $W$ and $L$ simultaneously until
CST finds the smallest s11 across the band. Optimization was
still running after two days and we manually stopped it.

We found that for these transmission lines and for $R=20mm$, s11 has strong dependence on $W$ and $L$. However, all s11 curves pass through an apparent fixed point at approximately
$f=580$MHz, $S_{11}=-12$dB.  From this result and from a manual
exploration of transmission line impedance we concluded that this
optimization step is essentially minimizing s11 by matching
the petal shape to the fixed balun parameters.

We introduced an additional degree of freedom by dividing the vertical
portion of the transmission line from one segment to two segments with
different impedances.  From among more than 60 sets of parameters
returned by CST, we picked $(W,L)=(138.5, 131.9)$mm which has the
smallest s11 across the band although it does not meet our
specifications and explore transmission line properties.  We held total length of
vertical transmission line fixed to ensure $\frac{1}{4}\lambda$
separation between radiating petals and reflective ground plane, and
varied characteristic impedances and the step location.

 With the upper trace width 3.5mm, length 92mm and lower trace
width 2.5mm, length 40mm, the result is dramatic. The fixed point is
removed and the s11 is below -15dB across the band except for
near 400MHz, where we just meet our requirement of -10dB.

We stopped our tuning procedure at this point.  Although a solution has been found which exceeds our
requirements, the system has not been optimized. Petal shape
parameters and transmission line parameters have been varied
separately but the full space of these parameters has not been
explored.  We can use this in future work to add additional
performance criteria to the design procedure.
\section{results}
\begin{figure}[!t]
\centering
\includegraphics[width=0.5\textwidth]{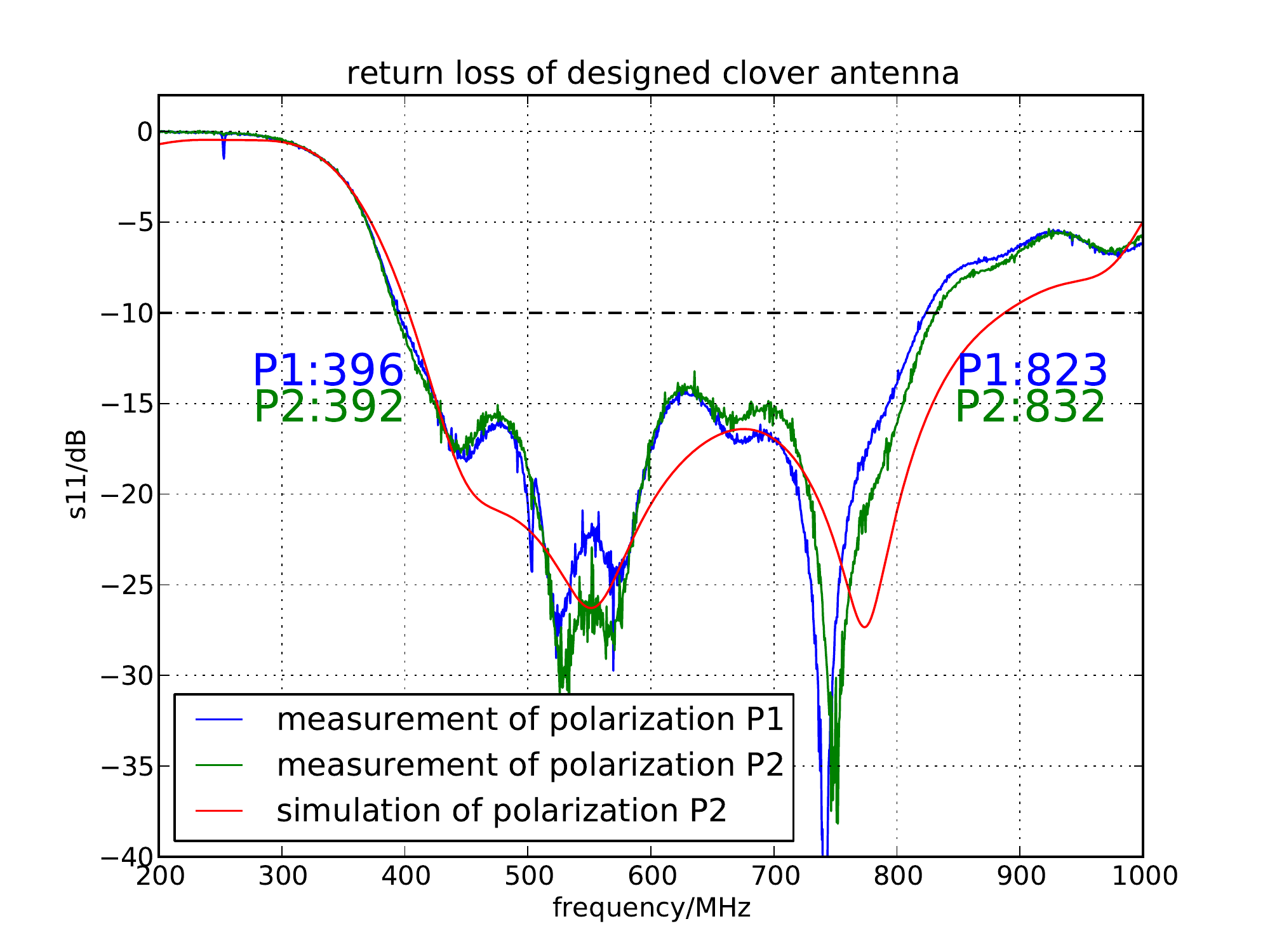}
\caption{The measured s11 spectrum for both linear
polarizations is plotted along with the CST simulation. Note the similarity between two polarzations.
This design exceeds the requirement of $S_{11}\leq -10$dB over the full band from 400 to 800 MHz.
} 
\end{figure}
\begin{figure}[!t]
\centering
\includegraphics[width=0.5\textwidth]{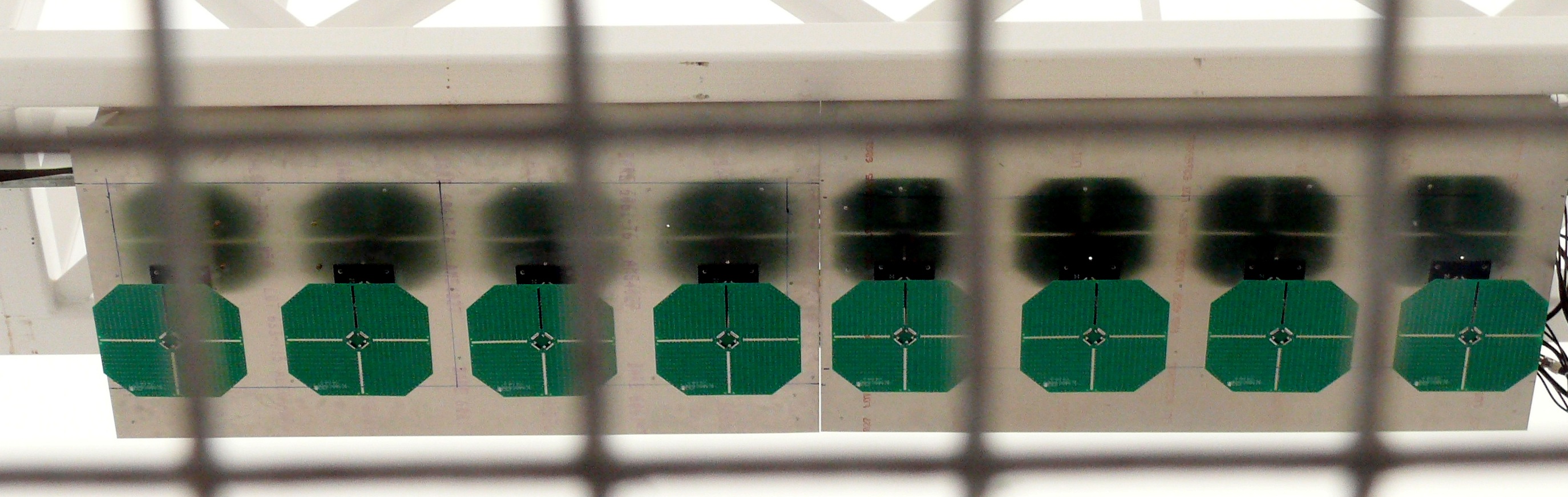}
\caption{A linear array of eight cloverleaf antennas installed at the
focal line of the CHIME Pathfinder at the Dominion Radio Astrophysical
Observatory in Penticton, BC, Canada.  The picture is taken through
the wire mesh reflective surface (mesh spacing 19 mm) illustrating a
photons
 view of the antennas and ground plane.  Notice that each feed
has an image-feed in the ground plane,  $\frac{1}{2}\lambda$ away at the
passband centre frequency.  Notice also the four slots cut to remove
dielectric material from the gaps between the petals.}
\end{figure}
Four petals of the chosen shape are built into one piece of double-sided PCB with FR4 as substrate to save cost. Vias connect the two copper surfaces to reduce material loss in FR4. The circuit boards are slotted to
remove FR4 in the gaps between petals because leaving FR4 in the gaps has a serious effect on both antenna impedance and material losses. Note that the resulting petal size and shape are compatible with 45
degree aimuthal rotation in an array. The s11 of an assembled feed is
shown in Fig. 2 for both polarizations and in comparison with
simulations. 
According to simulation, the beam pattern is smooth in both
the E-plane and the H-plane. HPBW varies within several degrees across the
band. The six PCB pieces of the cloverleaf antenna are
soldered together using a mechanical jig.  A photo of eight antennas in a
linear array installed on the CHIME pathfinder is shown in Fig.3.
\section{Acknowledgements}
CHIME is funded by the four partner institutions, grants from NSERC and the Canada Foundation for Innovation. MD acknowledges support from MITACS. DCW acknowledges support from Sydney University for this work.

\end{document}